\documentclass[12pt,graphicx,subfigure,axodraw]{article}
\usepackage{amsfonts}
\usepackage{amssymb}
\usepackage{epsfig}
\usepackage[usenames,dvipsnames]{color}

\newcommand{\beqn}{\begin{eqnarray}}
\newcommand{\eeqn}{\end{eqnarray}}
\newcommand{\be}{\begin{equation}}
\newcommand{\ee}{\end{equation}}

\newcommand{\te}{\theta}
\newcommand{\thb}{\bar\theta}

\def\st{Stueckelberg~}
\def\s1{$s_{\alpha}$}
\def\s2{$s_{\gamma}$}
\def\s3{$s_{\delta}$}
\def\c1{$c_{\alpha}$}
\def\c2{$c_{\gamma}$}
\def\c3{$c_{\delta}$}

\def\DDM{Dirac dark matter~}

\def\smnew{standard model~}

\begin{document}
\baselineskip 18pt

\thispagestyle{empty}

\vspace{0cm}

\begin{center}
{ {\bf  \Large
{
Multicomponent Dark Matter in  Supersymmetric Hidden Sector Extensions
}  } 
 
}
\vspace{1.0cm}

{\bf Daniel Feldman}\footnote{e-mail: djfeld@umich.edu}$^{,a}$,
{\bf Zuowei Liu}\footnote{e-mail: liu@max2.physics.sunysb.edu}$^{,b}$,
{\bf Pran Nath}\footnote{e-mail: nath@neu.edu}$^{,c}$,
{\bf and Gregory Peim}\footnote{e-mail: peim.g@neu.edu}$^{,c}$
\let\thefootnote\relax\footnotetext{ \\Preprint Numbers: MCTP-10-15, YITP-SB-10-08, NUB-3266}
\\
\vspace{.5cm}

{\it
$^{a}$Michigan Center for Theoretical Physics,
Ann Arbor, Michigan 48104, USA \\

$^{b}$C.N. Yang Institute for Theoretical Physics,
Stony Brook, New York, 11794, USA

$^{c}$Department of Physics, Northeastern University,
Boston, Massachusetts 02115, USA \\

}

\end{center}

\vspace{0.2cm}

\begin{center}
{\bf Abstract} \\
\end{center}
\vspace{0cm}
{Most analyses of dark matter within supersymmetry assume the  entire  cold dark
  matter  arising  only from weakly interacting neutralinos.}
We study a new class of models consisting of $U(1)^n$ hidden sector extensions of the minimal supersymmetric standard model 
that includes several stable particles, both fermionic and bosonic, which can be interpreted as constituents of dark matter. 
In one such class of models, dark matter is made up of both a Majorana dark matter particle, i.e.,  
a neutralino,  {and a Dirac fermion}  with the current relic density of dark 
matter as  given by WMAP being composed of the relic density of the two species. 
These models can explain the PAMELA positron data and are consistent 
with the antiproton flux data, as well as the photon data from FERMI-LAT.
Further, it is shown that such models can also simultaneously produce spin-independent cross sections 
which can be probed  in CDMS-II, XENON-100 and other ongoing dark matter  experiments. 
The  implications of the models  at the LHC  and at the next linear collider (NLC)  are also briefly discussed. 

\setcounter{footnote}{0}
\clearpage
{\section{Introduction \label{intro}}}
Recently several particle physics models have {been} constructed that connect the {\smnew{(SM)}} to  hidden sectors
and lead to massive narrow vector boson resonances {as well as} other signatures which can be detected at colliders 
{ \cite{kn12,FLNPRL,kctc}}. 
The connection to the hidden sector arises via mass mixings and kinetic  
mixings\cite{kn12,FLNPRL,kctc,Holdom,Dienes:1996zr,FLN07} and via higher dimensional operators.
{Models with the above forms of communication between the  sectors}
 also have important implications for dark matter {\cite{FKN,kctc,FLN07}} {{(for a review see \cite{white,Nath})}.}
  In this work we show that multicomponent dark matter can arise 
 {from}  $U(1)^n$ extensions of the {minimal supersymmetric standard model }
 (MSSM)  with {Abelian} hidden {sectors} which include hidden sector matter. 
 {Our} motivation stems in part from {the} results {of several} dark matter experiments that have {recently} appeared.
Thus the PAMELA Collaboration  \cite{PAM} has observed  
a positron excess improving previous results from HEAT and AMS experiments { \cite{heatams}}.
One possible explanation of such an excess is via the annihilation of dark matter in the
galaxy\cite{Feldman:2008xs}. Additionally, recent data from {CDMS-II} hints  at  the possibility of dark matter 
events above the background, and this will be explored  further by the upgraded 
XENON experiment  \cite{CDMSXENON,Aprile:2010um}. 

For a thermal relic, the PAMELA data and  {CDMS-II} data  taken together
at face value do raise a theoretical 
puzzle if indeed both signals arise  from the annihilation of cold dark matter. 
Thus most models {which aim to explain the PAMELA  positron excess} do not give a significant number of dark matter 
events in  the direct  detection experiments currently operating. {Conversely,} models which
{can} give {a} detectable signal in direct detection experiments {typically} do not explain the PAMELA data
without the use of enormous so-called boost factors. As we will show here, this can  {be}
circumvented in models where the dark matter has several components. Thus, motivated in part
by the recent cosmic anomalies  we develop supersymmetric models which 
contain minimally a hidden Abelian sector broken at the sub-TeV scale
where the mass generation of the hidden states involves nontrivial mixings 
with the field content of the electroweak sector of the minimal supersymmetric extension of the \smnew
 leading  to dark matter which can have several components  which can 
be both bosonic and fermionic.

More specifically, in this work we  go beyond the simple theoretical construction
 that thermal dark matter compatible with  WMAP  observations
 is composed of a single fundamental particle. 
 There is no overriding principle that requires such a restriction, and nonbaryonic dark matter (DM)
 may indeed be  constituted of several components, so in general one has
 $(\Omega h^2)_{DM} = \sum_{i} (\Omega h^2)_{DMi}$,
 where $i$ {refers} to the various species of dark particles  that can contribute to the total
 nonbaryonic $(\Omega h^2)_{DM}$. 
In fact we already know that neutrinos do contribute to dark matter although their contribution
is relatively small.  Thus   we propose here a new class
 of multicomponent  cold dark matter models in Abelian $U(1)$ extensions of MSSM which can
  simultaneously provide an explanation of the PAMELA 
and WMAP data  { through a Breit-Wigner enhancement \cite{Feldman:2008xs}}, 
while producing detectable  signals 
 for the direct searches for dark matter with CDMS/XENON
 and other dark matter experiments.
 
 A simultaneous satisfaction of  the PAMELA  
  positron excess and the satisfaction of  WMAP 
  relic density constraints can also occur if there is  a nonthermal mechanism 
 for the
  annihilation of dark matter with a wino lightest (R parity odd) supersymmetric particle (LSP) \cite{MoroiRandall,Kane,Acharya,Hisano,white,Nath}.
 However, a detectable
spin-independent cross  section in such a nonthermal  framework does require  
that a pure wino is supplemented by a suitable admixture of Higgsino content as in the analysis of  \cite{FKRN}
and in  \cite{Feldman:2009wv}{,} the later for a thermal relic.
  We remark that multiple  $U(1)$ factors and its influence on dark matter 
 have very recently been studied \cite{Feldman:2009wv,Arvanitaki}.
 We  also remark, some other works have recently looked at dark matter with more than 1 component
 \cite{Hur:2007ur}.  The models proposed {and analyzed} here are very different from these.

The outline of the rest of the paper is as follows: In Sec.(\ref{hidden}) we give a detailed description of the
two models one of  which is based on a $U(1)_X$ extension of the MSSM where $U(1)_X$ is a hidden
sector gauge group with  Dirac fermions in the hidden sector. This model allows for dark matter 
consisting of Dirac, Majorana, and spin zero {particles}. The second model is based on a 
$U(1)_X \times U(1)_C$ extension of MSSM, where $U(1)_C$ is a {gauged} leptophilic symmetry and
 $U(1)_X$, as before, is the hidden sector gauge group   which also contains Dirac particles  in the hidden
sector. 
This model too has Dirac, Majorana, and spin zero particles  as possible dark matter. 
In both {cases} we will {primarily} focus on the possibility that dark matter  consists  of Dirac  and Majorana 
particles, {and  we will not discuss in detail  the possibility of  dark matter with 
bosonic degrees of freedom.} 
In Sec.(\ref{relic}) we discuss the relic densities  in the two component models. In Sec.(\ref{pamela})
we give  an analysis of the positron, antiproton, and photon fluxes in the two models. 
In Sec.(\ref{events}) we  give an analysis of event rates  for the proposed models 
 for {CDMS-II} and for {XENON-100.}
 We give the analysis within the {framework of} supergravity grand {unified models}  \cite{Chamseddine:1982jx,Arnowitt:1992aq}  defined by the parameters $m_0, m_{1/2}, A_0, \tan\beta$, and sign$(\mu)$  with nonuniversalities  
 (NUSUGRA) 
  defined by $\delta_{1,2,3}$  {in the gaugino sector}
{so that} $U(1)_{{Y}}\times SU(2)_L\times SU(3)_C$ gaugino masses at the 
grand unified theory (GUT) scale are given by $\tilde m_{i} = m_{1/2} (1+\delta_i)$ {($i=1,2,3$)} (see, e.g., \cite{Corsetti:2000yq}
  and references {therein}).   
 {We} also discuss
the possible  new physics one might {observe} at the LHC (for a recent review see also  \cite{Nath}) and elsewhere
 for these models. Conclusions  are given in Sec.(\ref{conclusions}).

\section{Multicomponent Hidden Sector Models}{\label{hidden}}
\subsection{Multicomponent $U(1)_X$ model}{\label{model1}}
A  $U(1)_X$  extension of the minimal supersymmetric \smnew
involves the coupling of a Stueckelberg chiral multiplet
$S=(\rho+i \sigma,\chi_S,F_S)$ to vector supermultiplets $X,B$,
where $\rho$ is a real scalar and $\sigma$  is an axionic pseudoscalar.
Here $X$ is the $U(1)_X$ {vector} multiplet which is neutral with respect to the SM gauge group with components $X =(X_\mu, \lambda_X, D_X),$ 
and $B$ is the $U(1)_Y$  {vector} multiplet with components   {$(B_\mu, \lambda_B, D_B)$, {where}
{the components are written in the 
 Wess-Zumino gauge}. The {chiral  multiplet  $S$} transforms under both $U(1)_X$ and $U(1)_Y$ and
acts as the connector sector between the visible and the hidden sectors.  The total Lagrangian 
of the system is given by 
\beqn
{\cal L}= 
{\cal L}_{\rm MSSM} +{\cal L}_{ U(1)_X}+  {\cal L}_{\rm St}
\label{stext}
\eeqn
where ${\cal L}_{ U(1)_X}$ is the kinetic energy piece for the $X$ {vector} multiplet and 
${\cal L}_{\rm St}$ is the supersymmetric \st mixing between the $X$ and the $B$  {vector}  
{multiplets} so that 
 \cite{kn12,FKN} (see also 
  \cite{Anastasopoulos,Coriano,Feldman:2009wv})%
\beqn
{\cal L}_{\rm St} = \int d^2\te d^2\thb\ (M_1 X+ M_2 B+  S +\bar S )^2\ ,
\label{mass}
\eeqn
where $M_1$ and $M_2$ are {mass} parameters.  The Lagrangian of Eq.(\ref{stext}) is invariant under
the $U(1)_Y$ and $U(1)_X$  gauge transformations, i.e., under 
\beqn \label{stgauge}
\delta_X X= \zeta_X + \bar\zeta_X\ ,\quad \delta_X S =  - M_1 \zeta_X ,~~
\delta_Y B= \zeta_Y + \bar\zeta_Y\ ,\quad \delta_Y S =  - M_2 \zeta_Y,
\eeqn
{where $\zeta$ is an infinitesimal transformation chiral superfield.}
In component form we have for the \st sector with $U(1)_{X}\times U(1)_Y$
\beqn \label{stueck} 
{\cal L}_{\rm St} &=& - \frac{1}{2}(M_1 X_{\mu} +M_2 B_{\mu} +\partial_{\mu} \sigma)^2
- \frac{1}{2} (\partial_\mu \rho)^2 
- i \chi_S \sigma^{\mu} \partial_{\mu}\bar {\chi_S} + 2|F_S|^2 
\nonumber\\
&& 
\hspace{0cm}
+\rho(M_1D_X +M_2 D_B)
+{{\bar \chi}_S} (M_1\bar \lambda_X + M_2\bar \lambda_B) 
 + \chi_S (M_1 \lambda_X + M_2 \lambda_B) \ . 
\label{ls}
\eeqn
In addition, one may include 
a supersymmetric kinetic mixing term between the $U(1)_X$ and $U(1)_Y$ gauge fields
 \cite{FKN} 
leading to 
{${\cal L}= 
{\cal L}_{\rm MSSM} +{\cal L}_{ U(1)_X}+{\cal L}_{\rm KM} +  {\cal L}_{\rm St}$}, where 
 \beqn
{{\cal L}_{ U(1)_X}+{{\cal L}}_{\rm KM}} &=& - \frac{1}{4} {X}^{\mu\nu}
{X}_{\mu\nu}  -i  \lambda_{X} \sigma^{\mu} \partial_{\mu} \bar \lambda_{X}
+\frac{1}{2} D_{X}^2 \nonumber\\
&& - \frac{\delta}{2} {X}^{\mu\nu} B_{\mu\nu}  -i \delta( \lambda_{X}
\sigma^{\mu} \partial_{\mu} \bar \lambda_B + \lambda_B \sigma^{\mu}
\partial_{\mu} \bar \lambda_{X})
+ \delta D_B D_{X}\ . \eeqn
One can also add additional $D$ terms as in  {\cite{FKN}}.
{Both Stueckelberg and kinetic mixings of the gauge fields $U(1)_X$ and $U(1)_Y$ 
are constrained by the electroweak data\cite{FLNPRL}.}
As a consequence of the {mixings,} the extra gauge  boson of the hidden sector couples
with the standard model fermions and can become visible at  colliders.}
{The Lagrangian for matter interacting  with the $U(1)$  gauge fields is given by}   
\beqn
{\cal L}_{\rm matt} ~=~ \int d^2\te d^2\thb\, 
\sum_i \Big[ \bar \Phi_i e^{2g_Y Q_Y B+ 2g_X Q_X X} \Phi_i 
+ \bar \Phi_{{\rm hid},i} e^{2g_Y Q_Y B+ 2g_X Q_X X} \Phi_{{\rm hid},i}\Big] \ .
\label{7}
\eeqn
where the visible sector chiral superfields are denoted by $\Phi_i$ (quarks, squarks, leptons, sleptons, Higgs, and Higgsinos of the MSSM) and 
the hidden sector chiral superfields are denoted by $\Phi_{{\rm
hid},i}$.
In the above, $Q_Y$ is the hypercharge normalized so that
$Q = T_3+Q_Y$. As mentioned already, the SM matter fields do not
carry any charge under the hidden gauge group and vice versa, i.e. $Q_X \Phi_i
=0$ and $Q_{\rm SM} \Phi_{\rm hid}=0$.
The  minimal matter content of the hidden sector consists of a
left chiral multiplet $\Phi_{\rm hid}  =(\phi, f, F)$ and a charge  conjugate 
$\Phi_{\rm hid}^c=   (\phi', f', F')$ so that 
$\Phi_{\rm hid}$ and 
$\Phi_{\rm hid}^c$ have opposite 
$U(1)_X$ charges and form an  anomaly-free combination.
A mass  $M_{\psi}$  for the Dirac field $\psi$ arises from an additional term in the  
superpotential $W_{\psi}= M_{\psi}  \Phi \Phi^c$, where $\psi$ is  composed of $f$ and $f'$.
The scalar fields acquire soft masses of size $m_0$  from spontaneous breaking of supersymmetry 
by gravity mediation,
and in addition acquire a mass from the term in the superpotential so that 
\beqn
m_\phi^2= m_0^2 + M_{\psi}^2 =m_{\phi^{'}}^2.
\eeqn
After spontaneous breaking  of the electroweak symmetry  there would be mixing between the vector fields
$X_{\mu}, B_{\mu}, A_{3\mu}$, where $A_{3\mu}$ is the third component of the $SU(2)_L$ field 
{$A_{a\mu}$, $(a=1,2,3)$.} After diagonalization {$V^T=(X, B, A_3)$} can be expressed
in the terms of the mass eigenstates  $E^T=(Z', Z, \gamma)$ as  follows:
\beqn
{ V_i= O_{ij} E_{i}, ~~i,j=1-3, ~ ~~ E = (Z', Z, \gamma). }
\eeqn
The  neutral vector mass squared matrix is of the form given in Ref. [1] of  \cite{FLN07}.
Further,  the chiral fermions in the $S+\bar S$ multiplet together with the MSSM gauginos and 
{Higgsinos} will form a $6\times 6$ neutralino mass matrix whose eigenstates are six neutralino states
$\chi_a$,  $a=1-6$, where we assume that the set {$\chi_1^0\ldots \chi_4^0$} is the regular set of neutralinos
and $\chi_5^0,\chi_6^0$ are the two additional neutralinos that arise in the $U(1)_X$ extension. 
From the components  $\lambda_X, \bar \lambda_X$ and $\chi_S, \bar \chi_S$ that appear in Eq.(\ref{ls}),
we can form two Majorana fields $\Lambda_X$ and $\psi_S$ as follows:
 {
  \be
    \Lambda_{X}= \left( \begin{array}{c}  \lambda_{X\alpha} \\ \bar\lambda^{\dot \alpha }_X
 \end{array} \right),\\
~\psi_S= \left( \begin{array}{c}  \chi_{\alpha,S} \\ \bar\chi^{\dot \alpha}_{S}
 \end{array} \right).\\
  \label{lambdax}
\ee 
These components combine with the MSSM gauginos and Higgsinos to form a $6\times 6$ neutralino 
mass matrix whose eigenstates are the six neutralinos $\chi_a$, $(a=1-6)$. Thus $\Lambda_X$ and
$\psi_S$ can be expanded as linear combination of $\chi_a$, i.e.,  
\beqn
\Lambda_X= R_{1a} \chi_a, ~~{a=1-6}, ~\psi_S= R_{2a} \chi_a, ~~{a=1-6}
\eeqn
where $R$ is the unitary matrix which diagonalizes the $6\times 6$ neutralino mass matrix.  }
Further
the CP even Higgs sector is extended by the additional state $\rho$   \cite{kn12}.
The results outlined here give the following types of interactions:
\begin{enumerate}
\item
{There are interactions} of the Dirac fermion in the hidden sector with the \smnew particles via 
$Z$,$Z'$, $\gamma$ interactions. Thus, the \DDM can annihilate into \smnew particles via exchange of $Z$,$Z'$, $\gamma$ in  the early universe and in the galaxy. 
Depending on which of the two, Dirac or Majorana, is the heavier one may have Dirac particles
annihilating into Majoranas or the Majorana particles annihilating into Dirac fermions in the galaxy:
\be
\bar\psi \psi \to \chi \chi  ~~ {\rm or} ~~ \chi \chi \to  \bar\psi \psi ~.
\ee
\item
In addition to the above we have  fermion-neutralino-sfermion couplings
in the hidden sector  as given by Eq.(\ref{7}).  Thus interactions of the type 
$\bar \psi \chi_a \phi +{\rm h.c.}$,
etc. can produce decays such as $\phi \to \psi + \chi_a$ if they are kinematically allowed.  
 \item
{The scalar field} $\rho$ is CP even and mixes with the MSSM Higgs fields.  Through these  mixings $\rho$ has
 couplings  to the SM fermions and  through these couplings it can decay into the SM fermions.
 
\end{enumerate}
  It is instructive to list all the new particles in this $U(1)_X$ model as summarized below:
  \beqn
  {\rm New ~particles ~of~the} ~U(1)_X ~{\rm model}\nonumber\\
   {\rm {spin ~0}:} ~\rho, \phi, \phi', \nonumber\\
   {\rm spin} ~ \frac{1}{2}: \psi,  \chi_5^0, \chi_6^0,\nonumber\\
   {\rm spin} ~1: Z'.
   \eeqn
   We assume that the  lightest $R$-parity odd {particle}
     (LSP) is the least massive neutralino {($\chi^0=\chi_1^0 \equiv \chi$)} and resides in the visible sector and thus the masses of $\chi_5^0, \chi^0_6$ 
   are larger than the LSP $\chi^0$ mass, and consequently $\chi_5^0, \chi^0_6$ are unstable and decay
   into SM particles and $\chi^0$. {The bosons} $Z'$ and $\rho$ are unstable and decay into SM fermion pairs $f\bar f$ with
   the decay of the $\rho$ going dominantly through the process $\rho \to b\bar b$ or $\rho \to t\bar t$ if 
   $m_{\rho} > 2 m_t$. 
   The remaining three particles $\psi, \phi, \phi'$  are all milli charged and, consequently,
 at least one of them is stable. {If we assume} $m_{\phi}, m_{\phi'}> M_{\psi}$, at least  
 $\psi$ is always stable and the other two may or may not be stable.
  These along with the LSP give rise to various possible candidates for dark matter. 
Thus, depending on the relative masses of the Majorana, Dirac, and spin 0 particles
there are three possibilities for the constituents of dark matter as {outlined} below.
\\

\noindent
{\bf Two component dark matter: Majorana + Dirac}{.--} 
This model arises as follows: consider the case where 
$m_{\phi} > M_{\psi} + M_{\chi}$. In this case the {decays} 
$\phi, \phi'\rightarrow  \psi + \chi^0$,  {will occur}
and  $\phi, \phi'$ will be unstable. 
Thus $\psi$ is stable and so is $\chi$ under the assumption of R parity conservation.
Consequently,
   we will have two dark matter {particles; namely},
one a Majorana which is the LSP in the visible sector and the other a Dirac  in the
hidden sector. 
The Majorana and Dirac particles once created  will annihilate   as follows: 
\beqn
\psi + \bar \psi \to Z, Z', \gamma \to {\rm  SM + SM'},
\label{2.3}
\eeqn
\vspace{-1cm}
\beqn
{\chi+ \chi \to (s: Z',Z, h, H, A, \rho) ,({t/u}: \tilde f_{a},  \chi_{i} ,\chi^{\pm}_k)  \to {\rm SM + SM'}. }
\label{2.4}
\eeqn 
{where $s:$ and ${t/u}:$ refer to  $s$ and $t$ or $u$ channel exchanges.}
In addition to Eq.(\ref{2.4}) there are {coannihilation}   processes 
which contribute to the relic density. 
{Since both $\psi$ and $\chi$ are stable, the total relic density of dark matter will be the sum of the relic
densities for the two, the sum being  constrained by the WMAP data}.  These constraints are discussed further
in Sec.(\ref{relic}).\\

\noindent {\bf Three component dark matter:  Dirac and two spin 0 particles}{.--}
Suppose the mass of { $\chi$}   is larger  than the 
sum of the masses of the Dirac plus  the scalar $\phi$, i.e.,
$M_{\chi} > M_{\psi} + m_{\phi}$. 
In this case the decay $\chi \rightarrow  \phi+ \psi, \phi' +\psi$ will occur
and, consequently, $\chi$ is unstable. On the other hand, $\phi$,  $\phi'$ and $\psi$ are  stable 
since they cannot decay into anything else. Thus, here we have three dark matter particles:
{one  Dirac, and the other two spin 0.}  
Processes that {lead to the annihilations of}  these particles are those in Eq.(\ref{2.3}) for $\psi$,   and {also} for $\phi$ and
{$\phi'$,} they are similar to those in Eq.(\ref{2.3}), i.e., 
$\phi + \phi^*,  \phi' + {\phi'}^* \to \gamma, Z, Z' \to \rm SM + SM'$.
{In this three component dark matter model all the components 
reside in the hidden sector and thus their couplings to the standard  model particles 
are extra weak. Consequently, they will have very small spin-independent
cross sections in direct detection experiments. For this reason, this class of models is less 
preferred compared  to the two component model.}\\

 \noindent
 {\bf Four component dark matter: Majorana, Dirac, and two spin 0 particles}{.--} 
Finally, we consider the case when either of the following two situations occur:
(i) $M_{\chi} > M_{\psi},  m_\phi <   M_{\chi} < M_{\psi}+ m_\phi$, 
 (ii)    $M_{\chi} < m_\phi   < M_{\chi} + M_{\psi}$.
In these cases all four particles, one Majorana, one Dirac, and two spin 0 particles, are stable 
 and thus are possible dark matter candidates. 
 These particles will annihilate to the SM particles as in Eq.(\ref{2.3}), Eq.(\ref{2.4})
 and for $\phi$ and $\phi' $ via processes  in the three component 
 dark matter model as described above. 
This model is in many ways similar to the two component model and  like the two component model
 this model too should lead to detectable signals in  experiments for the direct detection of dark matter.

\subsection{Multicomponent Leptophilic  $U(1)_X\times U(1)_C$ model}
{We discuss now another model which contains two additional 
Abelian vector bosons where one of the extra bosons is leptophilic.} 
{Leptophilic} $Z'$s have  a long history  \cite{He:1991qd} and have been revisited
  \cite{Raidal} over the recent past in the context of dark matter.
Here we will consider a  
  $U(1)_X\times U(1)_C$ model where  the $U(1)_X$ as before is in the hidden sector,
 and $U(1)_C$ is a  {leptophilic} symmetry. 
 As in the $U(1)_X$ model, 
 we also assume that the hidden sector has  a pair of Dirac fermions $\psi$ and
$\bar \psi$ which are charged under $U(1)_X$ but  are neutral under the \smnew
gauge group and under $U(1)_C$.  Regarding $U(1)_C$ we assume it to be  $L_e-L_{\mu}$, i.e.,  
a difference of  family-lepton numbers,  which is anomaly free, and can be gauged.
The corresponding gauge field $C_{\mu}$ couples  only to $e,\mu$ {families} and nothing else.
The total Lagrangian in this case is 
\beqn
{\cal L}= 
{\cal L}_{\rm MSSM} + {\cal L}_{ U(1)^2}+  {\cal L}_{\rm St},
\eeqn
where ${\cal L}_{ U(1)^2}$ is the kinetic energy for the $X$ and $C$ {multiplets} and 
for ${\cal L}_{\rm St}$ we assume the following form: 

\beqn
{\cal L}_{\rm St} = \int d^2\te d^2\thb\ (M_1 C+ M^{\prime}_2 X +  M_3^{\prime} B+ S +\bar S )^2\ \nonumber\\
+ \int d^2\te d^2\thb\ (M^{\prime}_1  C+ M_2 X + M_3^{\prime\prime}B+  S' +\bar S' )^2,
\label{mass2}
\eeqn
where $C$ is the $U(1)_{L_e-L_{\mu}}$ vector multiplet with components $(C_\mu, \lambda_C, D_C)$
and $X$ and $B$ are the $U(1)_X$ and $U(1)_Y$  multiplets as discussed before.  
The gauge transformations under $U(1)_C$, $U(1)_X$, and $U(1)_Y$ are 
\beqn \label{stgauge}
\delta_C C = \zeta_C + \bar\zeta_C\ ,\quad \delta_C S = - M_1 \zeta_C\ ,
\quad \delta_C S' = - M_1' \zeta_C\,
 \nonumber\\ 
\delta_X X = \zeta_X + \bar \zeta_X\ ,  \quad \delta_X S = - M_2' \zeta_X\ ,
\quad \delta_X S' = - M_2 \zeta_X\ , 
 \nonumber\\ 
\delta_Y B = \zeta_Y + \bar \zeta_Y\ ,  \quad \delta_Y S = - M_3^{\prime} \zeta_Y\ ,
\quad \delta_Y S' = - M_3^{\prime\prime} \zeta_Y\ ,
\eeqn
where {$\zeta_C, \zeta_X, \zeta_Y$}, etc. are the {infinitesimal transformation chiral superfields.}
The quantities $M_1,\, M_2,\, M_1',\, M_2', M_3^{\prime},$ and  $ M_3^{\prime\prime}$ are the mass parameters. 
In the vector boson sector ${\cal L}_{\rm St}$  assumes the form 
\beqn
{\cal L}_{\rm St}= - \frac{1}{2} (M_1C_{\mu}+ M_2' X_{\mu} + M_3^{\prime} B_{\mu}+ \partial_{\mu} \sigma)^2 -  
\frac{1}{2}  (M_1' C_{\mu} + M_2 X_{\mu} +  M_3^{\prime \prime} B_{\mu} + \partial_{\mu} \sigma')^2.
\label{mass3}
\eeqn
The mass$^2$ matrix in the vector boson sector in the basis 
$(C^{\mu}, X^{\mu},  B^{\mu}, A^{3\mu})$ is given by 
 \be
\left( \begin{array}{cccc}
 M_1^2+  M^{\prime 2}_1  &   M_1 M^{\prime }_2 +  M^{\prime }_1 M_2&
 M_1M_3^{\prime}+ M_1^{\prime} M_3^{\prime \prime} & 0  \\
M_1 M^{\prime }_2 +  M^{\prime }_1 M_2 &     
 M_2^2 + M^{\prime 2}_2 & M_2^{\prime}M_3^{\prime}+ M_2 M_3^{\prime \prime}  & 0 \\
 M_1M_3^{\prime}+ M_1^{\prime} M_3^{\prime \prime} & M_2^{\prime}M_3^{\prime}+ M_2 M_3^{\prime \prime} &  M_3^{\prime 2} +M_3^{\prime \prime 2}+M_Y^2 &  -M_YM_W\\
0 & 0 & -M_YM_W & M_W^2 \\
 \end{array} \right)
 \label{vectormass}
\ee 
where $M_W {=g_2\cdot v/2}$ is the {$W$} boson mass and $M_Y =M_W \tan\theta_W {=g_Y\cdot v/2}$, 
and
where $\theta_W$ 
is the weak angle. {The dynamics of the model of Eq.(\ref{vectormass}) is  rather involved. 
We will focus, therefore, on a  simpler version of this more general case where we neglect the  mixings
with $B_{\mu}$, i.e., we set $M_3^{\prime}=M_3^{\prime\prime}=0$.  Inclusion of these coupling in the
analysis would not drastically change the analysis or the conclusions of this work as long as we keep
the mixing parameters $M_3^{\prime}/M_{1,2}, M_{3}^{\prime\prime}/M_{1,2}$ very small.}
After neglecting the mixings with $B_{\mu}$,
the mass$^2$ matrix is block diagonal and so we can  diagonalize  the top left hand corner 
 $2\times 2$ mass matrix  independent of the \smnew sector. 
 We are interested in the limit of small mixing between $U(1)_X$ and $U(1)_C$ and
 thus consider\footnote{{Note these mass terms $M_1,M_2$ are different than those considered in {Sec.} \ref{model1}}}  
 \beqn
M^{\prime}_1, M^{\prime}_2 ~ \ll~ M_1, M_2.
 \label{19}
  \eeqn
 In the above approximation the eigenvalues of this mass matrix are 
  \beqn
 M^2_{Z^{\prime}}\simeq M_2^2 + M_2^{\prime 2} - \Delta_{M^2}, ~~~
  M^2_{Z^{\prime\prime}}\simeq M_1^2 + M^{\prime 2}_1 + \Delta_{M^2}, \nonumber\\
  \Delta_{M^2} \simeq \frac{(M_1 M^{\prime}_2+ M^{\prime}_1 M_2)^2}{ (M_1^2+ M^{\prime 2}_1  -M_2^2-M^{\prime 2}_2 )}.
  \eeqn
 The corresponding mass eigenstates are $Z'$ and $Z''$, where  
\beqn
C_{\mu}=  \cos\theta_X Z''_{\mu} -\sin\theta_X Z'_{\mu}, 
~ X_{\mu} =\sin\theta_X Z''_{\mu} + \cos\theta_X Z'_{\mu},\nonumber\\
  \tan\theta_{X} \simeq \frac{ M_1  M^{\prime}_2+  M^{\prime}_1 M_2}{ M_1^2 +  M^{\prime 2}_1 -M_2^2 -M^{\prime 2}_2}. 
 \eeqn 
 Because of Eq.(\ref{19}) $\tan\theta_X \ll 1$.
   In the above, 
 the Dirac  fermions in the hidden sector have no couplings  with the photon and are electrically
 neutral. However, by a 
 {small mixing of  $X_{\mu}$  with  $B_{\mu}$  in ~Eq.(\ref{mass3})},  we can generate a 
 milli charge for the Dirac particles in the hidden sector  consistent with all electroweak data.

 We discuss now the {gaugino/chiral fermions in the extra $U(1)$ sectors which arise from the superfields}  $C, X, S+\bar S, S'+\bar S'$.
From the  {gaugino components} $\lambda_C, \bar \lambda_C$, $\lambda_X, \bar \lambda_X$, 
and from the   {chiral fermion components in the extra $U(1)$ sectors} 
 {$\chi_S, {\bar \chi}_S, \chi_{S^{\prime}}, {\bar \chi}_{S^{\prime}}$}, one can construct four component Majorana spinors two of which are exhibited in Eq.(\ref{lambdax}) and the remaining two are given by 
  \be
  \Lambda_{C}= \left( \begin{array}{c}  \lambda_{C\alpha} \\ \bar\lambda^{\dot \alpha }_C
 \end{array} \right),\\
 \psi_{S^{\prime}}= \left( \begin{array}{c}  \chi_{\alpha,S^{\prime}} \\ \bar\chi^{\dot \alpha }_{S^{\prime}}
 \end{array} \right).
  \label{e4}
\ee 
{ The neutralino mass matrix  in the 
{$[U(1)_X\times U(1)_C] \times [SU(3)_C\times SU(2)_L\times  U(1)_Y]$} model
takes a block diagonal form  }
{
\beqn
\left[
\begin{array}{c|c}
U(1)_X\times U(1)_C &  0_{4 \times 4} \\
  \rm  sector  &   \\
 \hline
    &  \\
 0_{4 \times 4} &  \rm MSSM \\
   & \rm sector
\end{array}
\right]_{8 \times 8}\ .
\eeqn
}
Thus, the Stueckelberg mass generation produces a
mass matrix in the hidden {gaugino/chiral fermion} sector which is decoupled from the neutralino
mass matrix in the visible sector.
Specifically in the 4 component notation the {gaugino/chiral fermion} 
mass matrix in the $U(1)_X\times U(1)_C$  sector  is given by 
 \be
L_{U(1)_X\times U(1)_C}^{mass}=
-\left( \begin{array}{c} \bar\psi_S \\ \bar \psi_{S'}   \\ \bar \Lambda_C   \\ \bar \Lambda_X
 \end{array} \right)^T
\left( \begin{array}{cccc}
0 & 0& M_1 & M^{\prime }_2\\
0 & 0&  M^{\prime }_1  & M_2\\
M_1 & M^{\prime }_1 & 0 & 0 \\
M^{\prime }_2 & M_2 & 0 &0 \\
 \end{array} \right)
 \left( \begin{array}{c}\psi_S \\  \psi_{S'}   \\  \Lambda_C   \\  \Lambda_X
 \end{array} \right).
 \label{e5}
\ee 
In the diagonalized basis we can label the extra neutralinos by $\chi_5^0,  \chi_6^0, \chi_7^0, \chi_8^0$.
Since the hidden sector and the neutralinos  of the visible sector are decoupled, the 
diagonalization of the neutralinos in the visible sector, i.e., of {$\chi_i^0$, $(i=1-4)$}  is not affected. 
{Further,} as for the case of the $U(1)_X$ model, it is instructive to list all the new particles in this 
$U(1)_X\times U(1)_C$ model
as summarized below:
\beqn
{\rm New ~particles ~of} ~U(1)_C\times U(1)_X {\rm model}\nonumber\\
{{\rm spin ~{0}:}} ~\rho, \rho', \phi, \phi', \nonumber\\
{\rm spin} ~\frac{1}{2}: \psi,  \chi_5^0, \chi_6^0, \chi_7^0, \chi_8^0,\nonumber\\
{\rm spin} ~1: Z', Z''.
\label{extra2}
\eeqn
We discuss now the stability of the new particles in this model. 
As before we assume 
that the mass of  $\phi$ (and of $\phi'$) is larger than the mass of $\psi$. Thus $\psi$ 
will be stable since it cannot decay into anything. If kinematically allowed the 
fields $\phi$ and $\phi'$ can decay  only via the process  $\phi, \phi'\to \psi+\chi^0$
as in the $U(1)_X$ model.
Of the remaining fields  obviously $Z'$ and $Z''$ are
unstable as they decay into $e\bar e, \mu\bar \mu, \nu_e\bar \nu_e, \nu_{\mu} \bar \nu_{\mu}$
as well as into $\psi\bar \psi$ depending on the mass of $\psi$. 
{As already noted,  a small milli charge can develop for the hidden sector matter via
small couplings of the $B_{\mu}$ and $X_{\mu}$ fields. The phenomenology of such models will
be very similar to the one we are discussing here.}
   
The extra neutralinos of Eq.(\ref{extra2}) 
can {also be}  all unstable. Thus $\Lambda_C$ couples  with 
leptons-sleptons ($e, \tilde e$ etc.) via coupling of the type {$\bar \Lambda_C e_L \tilde e_L^*$ },
etc. and after diagonalization of the {gaugino/chiral fermion} mass matrix all the $\chi_k^0$, $(k=5-8)$ will have
coupling with {leptons-sleptons} of the type indicated. Further,  two of the $\chi_k^0$ have roughly  a mass of size $M_1$ 
while the remaining two have roughly a  mass of size $M_2$. Thus, if $M_1, M_2 > m_{\chi^0}$,  which is 
what is assumed in this work, all the neutralinos of the hidden sector will be unstable and decay
into final states of the type $e\bar e \chi^0, \mu\bar \mu \chi^0$, etc. 
Regarding the field  {$\rho$,} there is an interaction of type
{\be 
 M_1g_C \rho (\tilde f^*Q_C^f \tilde f), ~~ f=e,\mu~. \ee}     
With this interaction $\rho$ will decay as follows:
$\rho\to \tilde f^* \tilde f\to f\bar f \chi^0\chi^0 (f=e,\mu)$ provided this process is kinematically allowed
which we assume is the case. A similar situation occurs for the case of $\rho'$. 
Additionally, if there is a mixing with $B_{\mu}$ in the \st sector then, as in the analysis of 
the $U(1)_X$ model, the fields $\rho$ and $\rho'$ will mix with the Higgs sector and can 
have decays of the type $\rho \to b\bar b$, $\rho'\to b\bar b$, etc ~.  
Thus, in the end we are 
left with  a similar  set of possibilities for dark matter as  in the $U(1)_X$ model, i.e.,
(i) a two component model with $\psi$ and $\chi^0$, (ii)  a  three component model with 
 $\psi$, $\phi$, $\phi'$, and  (iii) a four component model  with   $\psi$, $\phi$, $\phi'$,
and $\chi^0$.    
However, as in the $U(1)_X$ case we will focus on the two component 
model consisting of Dirac and Majorana dark  particles. 
 
We assume  $M_{Z^{\prime \prime}}^2 {\gg} M_{Z^{\prime }}^2$ and  that the annihilation of dark matter occurs close to the $Z'$ pole {for reasons that will become apparent shortly.} 
As a consequence, the annihilation of dark matter in the early universe
and in the galaxy is controlled by the $Z^{\prime}$ pole   {and the} effect of the $Z''$ pole on the analysis is essentially negligible. 
 The basic interaction of $C_{\mu}$ and of $X_{\mu}$ with matter is given by 
\beqn
{{\cal L}_{int}} = g_X Q_X \bar{\psi}  \gamma^{\mu} \psi X_{\mu}  + g_C Q_C^f \bar f \gamma^{\mu} f {C_{\mu}} 
\label{int}
\eeqn
{where $f$ runs over $e$ and $\mu$ families and where $Q_C^e=-Q^{\mu}_C$.}   
 In the mass  diagonal basis  the interaction of Eq.(\ref{int}) assumes the form  
\beqn
{{\cal L}_{int}} = 
(g_X Q_X \bar \psi \gamma^{\mu} \psi \cos\theta_{X} - g_C Q_C^f \bar f \gamma^{\mu} f \sin\theta_X)Z'_{\mu}\nonumber\\
+
(g_X Q_X \bar \psi \gamma^{\mu} \psi \sin\theta_{X} +  g_C Q_C^f \bar f \gamma^{\mu} f \cos\theta_X)Z^{\prime \prime}_{\mu}.
\label{28}
\eeqn
The  interaction of Eq.(\ref{28})
leads to the annihilation of $\psi \bar{\psi}$ into $e^+e^-$ and $\mu^+\mu^-$ via the 
$Z',Z^{\prime \prime}$ poles for which we assume a Breit-Wigner form. Thus, the $\psi \bar \psi \to f\bar f$ 
 annihilation cross section takes the form 
{ \beqn
\sigma_{\psi\bar \psi\to f\bar f} &=& a_{\psi} 
{\left|({s-M_{Z'}^2+i\Gamma_{Z'} M_{Z'}})^{-1} - 
({s-M_{Z^{\prime \prime}}^2+i\Gamma_{Z^{\prime \prime}} M_{Z^{\prime \prime}}})^{-1}
\right|}^2, \\
a_{\psi} &=& \frac{\beta_f (g_Xg_CQ_XQ^f_C\sin(2\theta_X))^2} {64 \pi s \beta_{\psi}}
\left[ s^2 (1+ \frac{{1}}{3} \beta_f^2\beta_{\psi}^2)
+ 4M_{\psi}^2 (s-2m_f^2) + {4} m_f^2 (s+ 2 M_{\psi}^2)\right],\nonumber\\
\eeqn}
where  $\beta_{f,\psi} = (1-4m_{f,\psi}^2/s)^{1/2}$.
{The relevant partial $Z'$ decay {widths are} given by}
{\beqn
\Gamma(Z'\to f \bar f ) &= &
(g_C{Q_C^f}\sin\theta_X)^2\frac{M_{Z'}}{12\pi},~~ f=e,\mu, \label{aa}\\
\Gamma(Z'\to \psi \bar \psi )&=&  ({g_XQ_X}\cos\theta_X)^2\frac{M_{Z'}}{12\pi}
\left(1+\frac{2M_{\psi}^2}{M^2_{Z^{\prime}}}\right)\left(1-\frac{4M_{\psi}^2}{M^2_{Z^{\prime}}}\right)^{1/2}
{ \Theta(M_{Z'} - 2 M_{\psi}),}
\label{aaa}
\eeqn}
{and similarly for the partial decay widths of the $Z^{\prime \prime}$ with 
$M_{Z^{\prime}} \to M_{Z^{\prime\prime}}$  and
{$-\sin\theta_X \to \cos\theta_X$} in Eq.(\ref{aa}) and 
{$\cos\theta_X \to \sin\theta_X$} in Eq.(\ref{aaa}).

{A constraint} on $g_C$ comes from the contribution of the $Z'$ and $Z^{\prime \prime}$ to $g_{\mu}-2$  \cite{Bennett:2004pv}.
Their exchange gives 
\beqn
\Delta (g_{\mu}-2) = \frac{ g_C^2 m_{\mu}^2}{ 24 \pi^2} \left [ \frac{\sin^2\theta_X}{M_{Z'}^2} + \frac{\cos^2\theta_X}
{M_{Z''}^2}\right].
\eeqn
{Using the current error \cite{Bennett:2004pv} of $\Delta (g_{\mu}-2)= 1.2\times 10^{-9}$ 
in the determination of $g_{\mu}-2$ and }
assuming $\theta_X$ is small, one finds the following constraint on $\alpha_C$:
\beqn
\alpha_C ~{\lesssim} ~0.001 \left(\frac{M_{Z^{\prime \prime}}}{300 {\rm~ GeV}}\right)^2 {~,}
\eeqn
{where $\alpha_C= g_C^2/4\pi$}. 
We   note that if the mixing angle $\theta_X$ is small, the decay width of $Z'\to f\bar f$ ($f=e,\mu$)
and of $Z^{\prime \prime}\to \psi \bar \psi$ will be  narrow while  the decay width of
$Z^{\prime \prime}\to f\bar f$ ($f=e,\mu$) and of $Z'\to \psi \bar \psi$ will be of normal size. However, 
 when $M_{\psi}\simeq M_{Z'}/2$ the $Z'$ decay width into $\psi \bar \psi$ will also be
 small due to the kinematic suppression  factor
 {$\left(1-\left[4M_{\psi}^2/{M^2_{Z^{\prime}}}\right]\right)^{1/2}$.  In this case we will have the total width of the
 $Z'$ to be rather narrow. Thus for annihilation near the Breit-Wigner pole we will have a  large
 enhancement of ${\langle\sigma v\rangle}$
  due to the narrowness of the $Z'$ { \cite{Feldman:2008xs}}.  It was shown in  the analysis of {Feldman-Liu-Nath} in  \cite{Feldman:2008xs}  
   that near the Breit-Wigner pole 
  such annihilations allow
 one to fit the relic density as well as allow an enhancement of ${\langle\sigma v\rangle}$ in the galaxy. 
 We note that while $Z'$ decay width is very small this is not necessarily 
  the case for $Z^{\prime \prime}$ which can decay into
 $e\bar e, \mu\bar \mu, \nu_e\bar \nu_e, \nu_{\mu}\bar \nu_{\mu}$ with normal strength. Thus neglecting the contribution of $Z^{\prime \prime}\to \psi\bar \psi$ 
 which is small due to the $\sin^2\theta_X \sim \epsilon^2$  suppression, one finds the total width of $Z^{\prime \prime}$
 to be 
 {
 $\Gamma_{Z^{\prime \prime}}\simeq \cos^2\theta_X \alpha_C M_{Z^{\prime \prime}}$.} 
 We will see in Sec.(\ref{pamela}) that the $\alpha_C$ needed in the analysis of the relic 
 density is {relatively small  compared to normal electroweak coupling} and, consequently, the width of $Z^{\prime \prime}$ 
 though significantly larger than the $Z'$ width is still relatively small 
 compared to
  what one might expect for a $Z'$ in a GUT model
  and certainly much smaller than the width
 for a $Z'$ arising as a Kaluza-Klein excitation in the compactification of an extra 
 dimension \cite{Nath:1999mw,Antoniadis:1999bq}.
Finally, the annihilation of the Dirac particles in the early universe goes by the processes 
\beqn
\psi\bar \psi \to Z', Z^{\prime \prime}\to e^+e^-, \mu^+\mu^-, \nu_e\bar \nu_e, \nu_{\mu} \bar \nu_{\mu},
\eeqn
 which is to be contrasted with the processes Eq.(\ref{2.3}) in the $U(1)_X$  model.

\section{Relic Density  in a Two Component Model}\label{relic}
Here we  discuss the relic density in models with two components.
{A general analysis requires solving the Boltzmann equations in a {Friedmann-Robertson-Walker} 
universe  \cite{Lee:1977ua,Goldberg:1983nd}, and includes coannihilations
 \cite{Griest:1990kh} and an accurate integration over pole regions.
As in the MSSM alone, one will generally encounter the $Z$ and Higgs poles  \cite{Nath:1992ty}
and these need to be treated with care. }
The {number} changing processes include 
{\beqn 
\psi \bar\psi \leftrightarrow  {\rm SM}~ {\rm SM}^{\prime},~~
\psi \bar\psi  \leftrightarrow  \chi \chi,~~
\chi \chi  \leftrightarrow { {\rm SM}~{\rm SM}^{\prime}.  }  
\eeqn 
}
{Note that the process  $\bar\psi \chi \leftrightarrow  {\rm SM}~ {\rm SM}^{\prime}$ is not allowed 
since $\bar\psi\chi$ connect only to $\phi$ and $\phi'$, neither of which can connect to the standard model
particles.}
{For the simplest two component model with dark matter particles $\psi,\chi,$ 
with the assumption that $M_{\psi} > M_{\chi}$
the only relevant processes in the annihilation of {$\psi\bar \psi$ are} 
{$\psi\bar \psi \to f\bar f,\chi \chi$}
final states. 
Since ${\psi}$ is heavier than $\chi$ its freeze-out occurs earlier {(at a higher $T$) }than for
$\chi$. 
Thus, the Boltzmann equations for $n_{\psi}$ (which includes fermions and antifermions)
and for $n_{\chi}$
{for the $U(1)_X$ and for the   $U(1)_X  \times  U(1)_C$ two component 
models are given by}
\be
\frac{dn_{\psi}}{dt} = -3H n_{\psi} - \frac{1}{2}{\langle\sigma v\rangle}_{\psi\bar\psi} (n^2_{\psi}- n_{\psi,{\rm eq}}^{2}), \label{rpsi}
\ee
\be
 \frac{dn_{\chi}}{dt} = -3H n_{\chi} - {\langle\sigma v\rangle}_{\chi\chi} ( n^2_{\chi} -    n^{2}_{\chi,{\rm eq}})
+\frac{1}{2}{\langle\sigma v\rangle}_{\psi\bar \psi \to \chi \chi} (n_{\psi}^2 -n_{\psi,{\rm eq}}^{2}).
\label{rchi}
\ee
Here ${\langle\sigma v\rangle}_{\psi \bar{\psi}}$ refers to $\psi\bar \psi \to f\bar f, {\chi \chi}$, 
and ${\langle\sigma v\rangle}_{\chi\chi}$ stands for  ${\langle\sigma v\rangle}_{\chi\chi\to {\rm \rm SM~ SM'}}$.
For the spin averaged cross section for the Dirac case, the extra factor of $1/2$ is 
to account for the fact that we are dealing with a Dirac fermion. The number densities are
 $n_{\psi},n_{\chi}$ and  $n_{\psi,\rm eq},n_{\chi, \rm eq}$ are their values at equilibrium, i.e., 
$
n_{(\psi,\chi),\rm eq} \simeq  g_{(\psi,\chi)} (M_{(\psi,\chi)} T)/2\pi)^{3/2} {\rm exp}({-\frac{M_{({\psi,\chi})}}{T}})
$,
where   $g_{\psi} =4$ and $g_{\chi} =2$.
{Since the {two} dark matter particles are sub-TeV in mass, they will freeze-out 
at temperatures that are not drastically different. 
{One can solve the Boltzmann equation for $\psi$ with the appropriate boundary conditions 
to compute the {freeze-out} temperature $T_f^{\psi}$  and the relic density {of $\psi$}  at 
the current temperatures.
To compute the freeze-out temperature $T_f^{\chi}$ for the particles $\chi$, one uses solutions for 
 $n_{\psi}$ as computed from the Boltzmann equation for $\psi$  as input in 
the Boltzmann equation for $\chi$  keeping in mind that $n_{\psi,\rm eq}$ in the $\chi$ Boltzmann equation can
 be neglected since  we are  below the freeze-out temperature for $\psi$. It is difficult to get a closed form
 solution of Eq.(\ref{rchi}) for $n_{\chi}$ and thus in general the analysis must be done  numerically for
$\Omega_{\chi} h^2$. However, it turns out that for both the $U(1)_X$ and the $U(1)_X\times U(1)_C$ 
models the contribution of the term proportional to $n_{\psi}^2$ in Eq.(\ref{rchi})
is rather suppressed and it is a good approximation to neglect this term  for both models. In this case,
one has
{
\beqn
 (\Omega h^2)_{\rm WMAP} =   (\Omega_{\psi} h^2)_0 + (\Omega_{\chi}  h^2)_0  \simeq  \frac{C_{\psi}}{J_0^{\psi}} +\frac{C_{\chi}}{J_0^{\chi}},
 \eeqn
 where 
 \beqn
C_{\chi} \simeq  \frac{1.07 \times 10^{9}~ {\rm GeV^{-1}}}{ \sqrt{g^{*}({\chi})}M_{\rm pl}}~,~~~
C_{\psi} \simeq  2\times \frac{1.07 \times 10^{9}~ {\rm GeV^{-1}}}{ \sqrt{g^{*}({\psi})}M_{\rm pl}},\\
J_0^{\chi} = \int_0^{x_f^{\chi}}  {\langle\sigma v\rangle}_{\chi\chi} ~dx~, ~~~~~
 J_0^{\psi} = \int_0^{x_f^{\psi}}  {\langle\sigma v\rangle}_{\psi {\bar{\psi}}} ~dx~,
\label{r3}
\eeqn
and}
where {$g^{*}({\psi,\chi})$ denotes}  the effective degrees of freedom at the freeze-out of $\psi,\chi$
respectively.
The analysis leading to Eqs.(\ref{rpsi},\ref{rchi}) is easily extended to include {coannihilations}.} The analysis
can easily be reversed if the Majorana is heavier than the Dirac.
{Denoting $\rho_{\odot,\psi}$, $\rho_{\odot, \chi}$ 
as the local density of each dark matter kind in the halo, one can assume 
\beqn
\rho_{\odot,\psi}/\rho_{\odot,\chi}~ \sim ~ (\Omega_{\psi}  h^2)_0/  (\Omega_{\chi} h^2)_0.
\label{46}
\eeqn
However, the ratios need not be {the} same. 
The local halo densities are also constrained 
such that  $\rho_{\odot,\psi}+ \rho_{\odot, \chi} 
= \rho_{\odot,\rm total} \simeq (0.35-0.45) \rm GeV  cm^{-3}$.
For the calculation near the $Z^{\prime}$ pole, we use the analysis of of  \cite{FLN07}
which follows the techniques of  \cite{Nath:1992ty}. {Indeed the analytic {techniques} developed in   \cite{Nath:1992ty,FLN07}
have been cross-checked with  independent codes.}
For the $U(1)_X$ model, the decay branching ratios are substantially less  hadronic and more leptonic  
 than for the annihilations  via the $Z$ boson exchange  \cite{Feldman:2006wb}. 
 For the $U(1)_X\times U(1)_C$ model the decays of the  $Z', Z^{\prime \prime}$ are purely leptonic. 
 These leptophilic decay patterns for the extra $Z's$ help to explain the PAMELA  positron excess without 
 recourse to large {\it{ad hoc}} boost factors.

\section{Positron, Antiproton, and Photon fluxes in the $U(1)_X$ and $U(1)_X\times U(1)_C$ Models}\label{pamela}
{An excess of positrons,  antiprotons and photons over  the cosmic  background is a possible 
indicator for  annihilating dark matter in the galaxy as was pointed out early on  \cite{Silk}.}
 {In our {multicomponent} supersymmetric models} contributions from the Majorana component to the fluxes are 
negligible {(suppressed by {an} order of magnitude or more)} and essentially the entire effect arises from the Dirac component.}
{The positron} flux $\Phi_{e^{+}}$ arising from the annihilation {of} 
dark matter (DM) particles  is 
  \cite{Delahaye:2007fr,Cirelli:2008id}
\be
\Phi_{e^{+}} = \frac{\eta v_{e^{+}} }{4 \pi b(E)}\frac {\rho^2_\odot}{M^2_{\rm DM}} \int^{  M_{\rm DM}  }_{  E }    \sum_f {{\langle\sigma v\rangle}_{ f,\rm halo}\frac{d N^f_{e^{+}}}{d E'}}~ B_{\bar e}~ {\cal I }_{(E,E')}d E'.
\label{1}
\ee
{Here $M_{\rm DM}$ is the mass of the dark matter particle.}
{In the above $\eta =(1/2,1/4)$ for (Majorana, Dirac) cases, respectively,} 
 $B_{\bar e}$ is a boost factor which may arise as a consequence of dark matter substructure, or a local clump.
{ Recent N-body simulations show that it is unlikely that large dark matter clumps exist within the halo of our
galaxy \cite{Brun:2009aj} and thus the use of large clump factors  in flux analyses  appear unreasonable} \cite{Kamionkowski:2010mi}. 

 The other parameters that 
  enter Eq.(\ref{1}) are as follows: 
 ${\cal I }_{(E,E')}$ is the  halo function and we   parametrize it in some of the  standard forms
adopted in the  literature  with the appropriate diffusion models  \cite{Delahaye:2007fr} using the
standard profiles
 \cite{Navarro:1996gjMoore:1999gc}.  The positron velocity 
  is  $v_{e^{+}} \sim c$,  and the {energy} loss function $b(E)$ has the form  
$b(E)  = E_0 (E/E_0)^2 /\tau_E$, where  $\tau_E  \sim (1-2) 10^{16} [\rm s]$, with $E$ in [GeV] and $E_0 \equiv 1 \rm GeV$.
{We use the GALPROP background estimate of  \cite{Moskalenko:1997gh} fit in  \cite{Baltz:1998xv}; modifications of the 
background estimates require either smaller or larger mass splitting at the pole which can range from {the} order a GeV to the order
of tens of MeV, depending on the assumed astrophysical background and the level of clumpiness of the signal. }
Further, in Eq.(\ref{1})  ${\langle\sigma v\rangle}_{\rm halo} $ is the velocity averaged cross section
in the {\em halo  of the galaxy}. {We note}  that  
${\langle\sigma v\rangle}_{\rm halo}$ may be significantly different than
 ${\langle\sigma v\rangle}_{x_f}$ at the {epoch of freeze-out}. Thus as emphasized in 
  \cite{Feldman:2008xs}
  a replacement of  ${\langle\sigma v\rangle}_{\rm halo}$ by 
 ${\langle\sigma v\rangle}_{\rm freeze-out}$, as is often done,  is inaccurate and can lead to significant errors in the
positron flux computation. {This stems from the fact that the relic density in previous works is often approximated
by pulling out   ${\langle\sigma v\rangle}$ from the integral between $T_{\rm freeze-out}$ and the current temperature}. 
{In general,
full integration must be taken into account in the vicinity of a pole for an accurate calculation  \cite{Nath:1992ty}  or when the dark matter {coannihilates}  \cite{Griest:1990kh}.
Both of these cases often arise in various parts of the parameter space of dark matter models. } 

\begin{figure}[h!]
\centering
 \includegraphics[width=11cm,height=8cm]{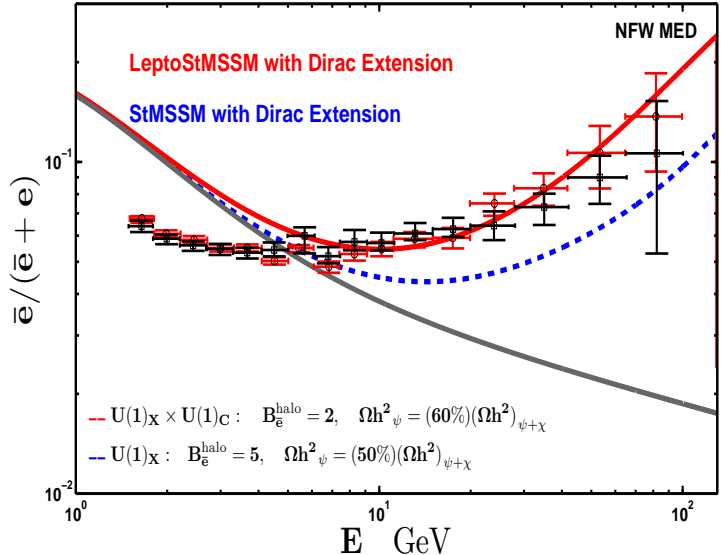}
\caption{Breit-Wigner enhancement  \cite{Feldman:2008xs} and the PAMELA positron excess.
The analysis presented here is given in the two component Dirac-Majorana 
$U(1)_X$  and  $U(1)_X\times U(1)_C$  models assuming 
a relic density decomposition of the  Dirac and Majorana as given in the figure.
The dominant contribution to the positron flux
comes from the annihilation of the Dirac particles in the galaxy. 
}
\label{positfig}
\end{figure}
In the numerical analysis
we fix $\rho_{\odot, \chi} = 0.18 \rm ~GeV/cm^3$ and take $\rho_{\odot, \psi} \in ( 0.18-0.25) ~\rm GeV/cm^3$.
{We allow the neutralino relic density to lie in the range}  {$(0.035,0.065)$} and find relatively good fits to the WMAP and PAMELA data{
for both the $U(1)_X$  and the $U(1)_X \times U(1)_C$ model with total relic density 
in the range ($\Omega_{\chi}  h^2)_0 +(\Omega_{\psi} h^2)_0 = (0.08, 0.12)$.}
The analysis of  Fig.(\ref{positfig}) shows  the PAMELA data and the positron flux ratio in the  $U(1)_X$ model
and in the $U(1)_X\times U(1)_C$
model consistent with assumed densities discussed above.}
 In this fit the dominant contribution comes from the annihilation 
$\psi \bar{\psi}\to Z'\to e^+e^-$ and  {only small} boost (clump) factors are used here, i.e., 
{$B_{\bar e}=(2-5)$}. 
Thus, the Breit-Wigner enhancement  \cite{Feldman:2008xs}
 plays an important role in achieving a simultaneous fit to the
 relic density and to the positron excess. Indeed, the annihilation  near a Breit-Wigner pole  gives a significant enhancement to the
annihilation cross section of \DDM  in the galaxy obviating the necessity of 
using large boost factors. 
At present, we are guided by the WMAP and PAMELA data on the mass splittings between the Dirac component of dark matter
and the vector boson mass. Generically the required splitting is 
$2 M_{\psi} - M_{Z'} \sim (10^2 - 10^3)~\rm MeV$ depending on the leptophilic nature
of the models {and narrowness of the resonance}. In Fig.(\ref{positfig}) the $U(1)_X\times U(1)_C$ has an 80 MeV mass splitting
 and {the} $U(1)_X$ model has 1300 MeV mass splitting. 
We note that since the $U(1)_X$ model is less leptophilic than the $U(1)_X\times U(1)_C$ model,
we have used from the parameters of the model a smaller annihilation cross section for the $U(1)_X$ model 
 than for the  $U(1)_X\times U(1)_C$ model throughout 
the analysis as indicated by the positron ratio in the figure. 
This is motivated by the constraints from the {antiproton} fluxes.
Thus, the $\bar p $ flux takes the form}
\beqn \Phi_{\bar p}(T) =   \frac{\eta v_{\bar p}}{4\pi}
\,  \frac{\rho^2_\odot}{M^2_{\rm DM}} \, B_{\bar p} \,R(T) \, 
\sum_f {\langle\sigma v\rangle}_{f,\rm halo}  \, \frac{dN^f_{\bar p}}{dT} \eeqn
where $T$ is the kinetic energy, and  $R(T)$ has been fit as in Ref. \cite{Cirelli:2008id} for
various profile/diffusion models and background estimates have been obtained in  \cite{Maurin:2001sj}.
The {antiproton} flux observed at the top of the atmosphere including
solar modulation can be accounted for by replacing
{$\Phi_{\bar{p}}(T)\to\Phi_{\bar{p}}\left(T+|Z|\phi_F \right)$} 
and including a kinetic
energy correction ratio. We take the Fisk potential
$\phi_F$ as 500~MV.
\begin{figure}[h!]
\centering
\includegraphics[width=10cm,height=8cm]{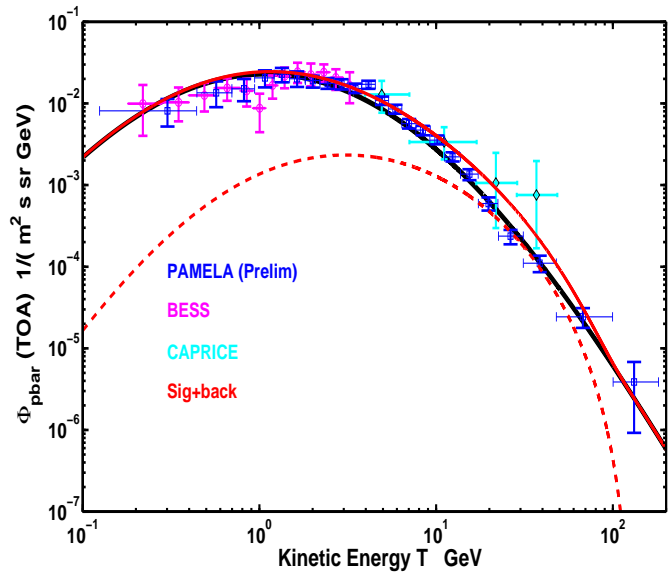}
\caption{ Antiproton flux 
in the $U(1)_X$ (signal and signal plus background). The analysis is done with the NFW median model while the minimum diffusion model 
is unconstrained by the $\bar p $ data  \cite{picozza} and is not shown. For the $U(1)_X\times U(1)_C$, leptophilic model the
$\bar p $ flux from the {Dirac component of} dark matter does not contribute to a signal. No boost factor from clumping is taken. 
{In both models the Majorana flux is highly suppressed  relative to the Dirac flux, and is thus not shown separately.}
Annihilation cross {sections} and local dark matter {densities} are those used in {Fig.}(\ref{positfig}).}
\label{pbarfig}
\end{figure}
In Fig.(\ref{pbarfig}) {we 
 exhibit}  an analysis of the antiproton flux (signal plus background) 
in the $U(1)_X$ model. The analysis is done with the {Navarro-Frenk-White (NFW)} median model while the minimum diffusion model 
is unconstrained by the $\bar p$ data and is not shown. {The PAMELA data exhibited in Fig.(\ref{pbarfig})
is taken from \cite{picozza}.} 
{For the $U(1)_X$ model the anti-proton flux overshoots a little bit beyond {$E= 10 ~\rm GeV$}
 but still {lies} within the limits of acceptability.
 For the  $U(1)_X\times U(1)_C$ model, the $Z'$ and $Z^{\prime \prime}$ 
 are both leptophilic 
and there are no annihilations of $\psi \bar{\psi}$ into $q\bar q$. Because of the absence of $q\bar q$
final states in the annihilation, there is no contribution to the {antiproton} flux from the annihilation
 of the {Dirac component of dark matter, thus the prediction of the model is not observable above the background.}
\begin{figure}[h!]
\centering
\includegraphics[width=10cm,height=8cm]{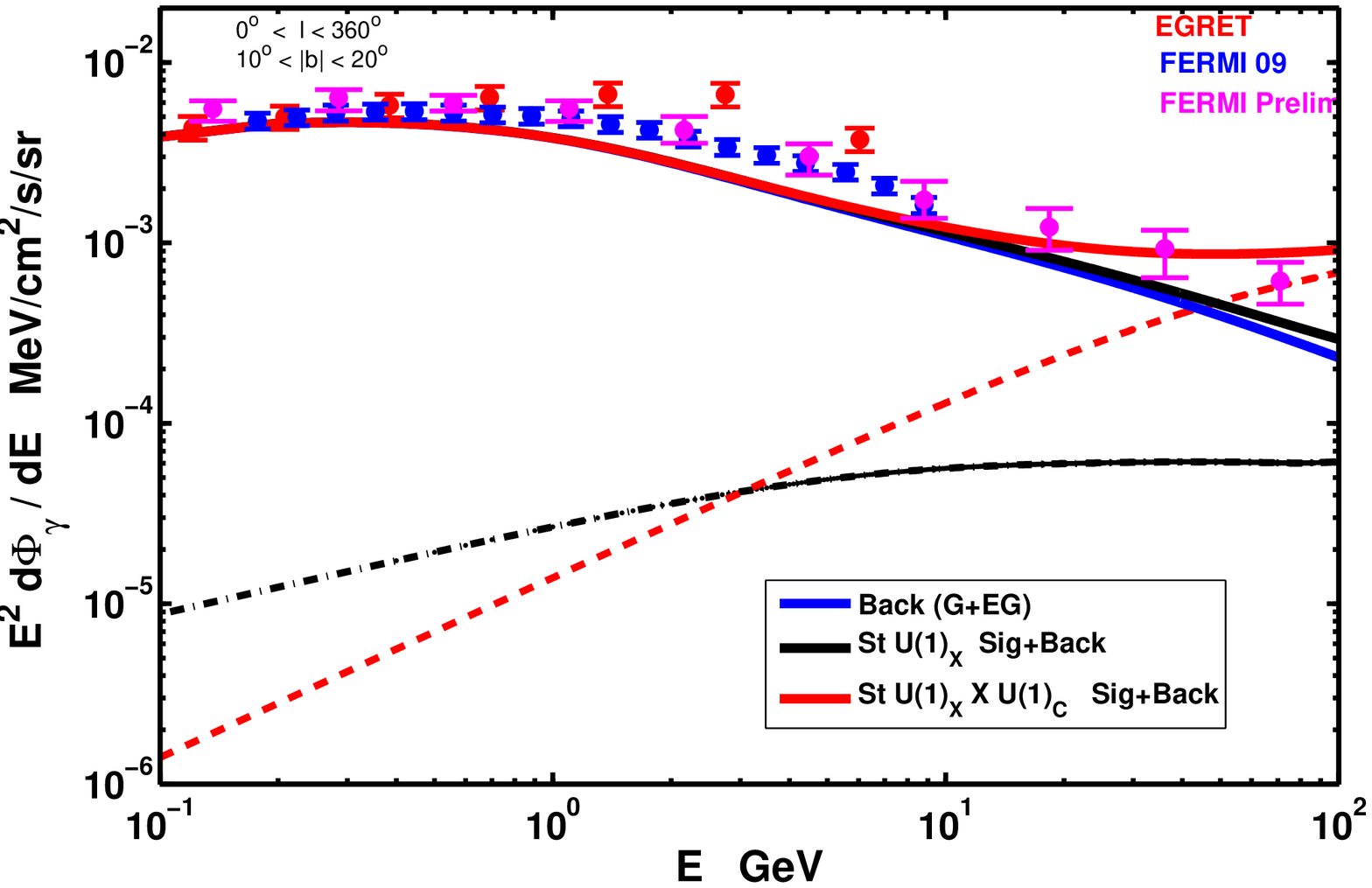}
\caption{Photon flux in the $U(1)_X$  and  $U(1)_X\times U(1)_C$ models. The photon flux for $U(1)_X$ includes {contributions}
from the quarks and taus and {bremsstrahlung}, 
while {the photon flux}
for the $U(1)_X\times U(1)_C$ model is highly suppressed at low energies and peaks at larger energies
from the {bremsstrahlung}. Annihilation cross {sections} and local dark matter {densities} are those 
{as}  in {Figs.}(\ref{positfig},\ref{pbarfig}).}

\label{photonfig}
\end{figure}
 Finally, we look at the photon flux. 
 In the  angular region $\Delta \Omega$ the (differential) photon flux (sometime denoted $d\Phi_{\gamma}/{dE}$) is given by 
\beqn
\Phi_{\gamma}   & = & \frac{\eta}{4 \pi}\frac{r_{\odot}\rho_{\odot}^2}{ M_{\rm DM}^2} \sum_{f}
{\langle\sigma v\rangle}_{f,\rm halo} \frac{dN_{\gamma}^f}{dE} \bar{J}\Delta \Omega, ~~~~~
\bar{J} = \frac{1}{\Delta \Omega}\int_{\Delta \Omega} \int_{los} \frac{ds}{r_{\odot}} 
\left( \frac{\rho(r(s,\psi))}{\rho_{\odot}}\right)^2 .
\eeqn
For the $U(1)_X$ model there are three contributions to the
 photon flux. These arise from the $q\bar q$, $\tau\bar \tau$, and from bremsstrahlung  
 {(see {i.e.}  \cite{Birkedal,Bergstrom:2005ss})}. For the
 $U(1)_X\times U(1)_C$ model since the $Z'$ and $Z^{\prime \prime}$ are leptophilic with allowed final states
 being only in the first two generations of leptons, there are no final states of the type $q\bar q$
 and {$\tau\bar \tau$} {for the Dirac component.}  
 However, there is an emission of continuum radiation  {for the Dirac component} 
 {(see e.g.,  \cite{Fornengo:2004kj} using {PYTHIA}  \cite{Sjostrand:2006za})}
which arises because $\psi \bar{\psi}$ annihilate
into $e\bar e$ and $\mu\bar \mu$ and there is an associated photon continuum radiation from bremsstrahlung.

{In Fig.(\ref{photonfig}) we give an  analysis of the continuum photon flux}
 {in the angular region where
the integral over the line of sight
is rather insensitive to the details of the dark 
matter distribution { \cite{Meade:2009iu,Cirelli:2009vg}
(for a recent analysis with focus on the galactic center see  \cite{Crocker:2010gy}).
The analysis is given for both $U(1)_X$ and  $U(1)_X\times U(1)_C$ models using 
 an {isothermal} profile. 
The continuum photon flux for the $U(1)_X$ model arises mostly from
 $q\bar q$ and $\tau\bar \tau$ at low energies while the 
final state {radiation, i.e.  {$e^+e^- \gamma$ and  $\mu^+ \mu^- \gamma$} takes over at high energies where $E_{\gamma}/M_{\psi} \to 1$}.  Also shown is the 
EGRET~ \cite{Hunger:1997we} data and 
the more recent FERMI-LAT data~ \cite{winer} as well as the background
flux in 10-20 region as estimated in the GALPROP {analysis} of  \cite{Reimer}. For the $U(1)_X\times U(1)_C$ model  
 the total photon flux {is the suppressed contribution from the Majorana} and  the
{dominant Dirac source arising  from the bremsstrahlung  }
 from the {final states} $e^+e^- \gamma$ and from $\mu^+ \mu^- \gamma$. }
One finds that the continuum spectrum {with bremsstrahlung} is in accord with the current 
{experimental} data. 
Regarding the monochromatic photon radiation from the annihilation of dark matter, it is suppressed
by $\epsilon^2  \lesssim 10^{-4}$ for the $U(1)_X$ model and the prediction for this model 
is far below the current experimental limits. 
For the $U(1)_X\times U(1)_C$ model there is no coupling of the hidden sector
  Dirac particles to the photon {if the mixing with the hypercharge gauge boson vanishes}.  
Thus,  at the tree level there {would} be no emission of  monochromatic radiation in the annihilation of dark matter
in this model. {In the case of a small or {nonvanishing} mixing with the hypercharge  as in the $U(1)_X$ model,
 this emission is also {suppressed.} }

  \section{ CDMS-II and  XENON }{\label{events}}
  {
The  {CDMS-II} results  mentioned  {in Sec.(\ref{intro})}} raise the possibility that 1-2 dark matter 
  events may have been seen in the  {CDMS-II} detector, and this possibility has led to a significant
  theoretical activity \cite{flncdms}. {Many analyses within supersymmetry assume the 
  supersymmetric cold dark
  matter is entirely composed of neutralinos.}  We {analyze} the event rates  in {CDMS-II}
  and in {XENON} detectors for the case when roughly   
  only half of the dark matter is constituted of neutralinos, a situation which holds for 
     both the  $U(1)_X$ and the $U(1)_X\times U(1)_C$ models.
  In the  analysis we use  {MICROMEGAS} \cite{Belanger:2008sj} and   
   impose the electroweak symmetry breaking constraints as well as all the current experimental
   constraints such as on $g-2$, flavor changing neutral currents, i.e.,  $b\to s\gamma$ and $B_s\to \mu^+\mu^-$
   branching ratio {constraints,} {(see i.e.} \cite{Chen:2009cw}, \cite{Cassel:2010px} and  \cite{Gomez:2010ga} for  
   recent analyses)
 and require that about half the relic density as given by  WMAP be given by neutralinos.}
   
\begin{figure}[h!]
\centering
   \includegraphics[width = 12cm, height = 10 cm]{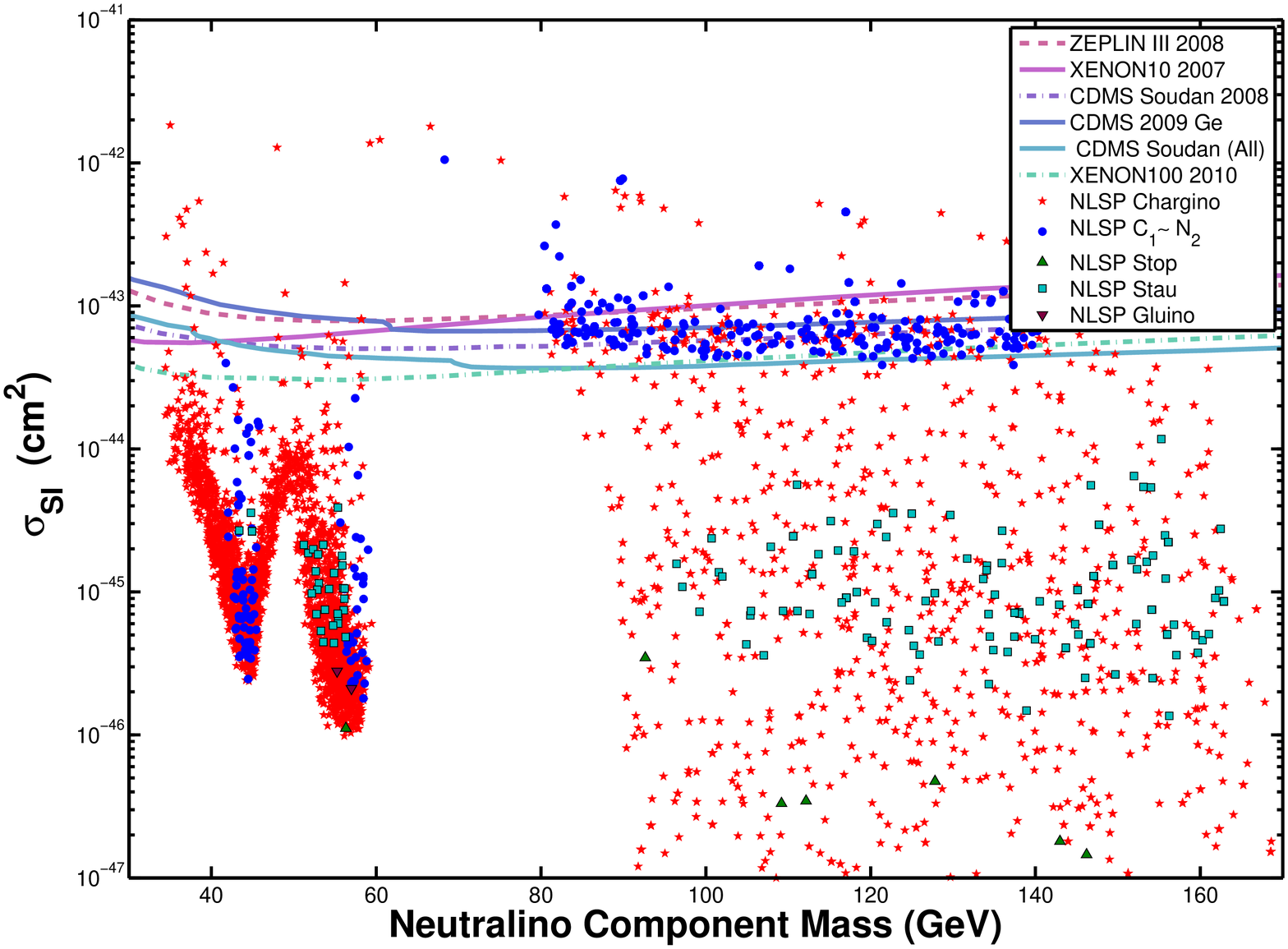}
\caption{An analysis of the spin-independent cross section for the parameter space
of supergravity models with nonuniversalities in the gaugino sector.  The  
{analysis given above is} valid for both the  
$U(1)_X$ and $U(1)_X\times U(1)_C$  models {with a neutralino component and a Dirac component of dark matter}.
Models are labeled by the NLSP, which under the constraints
of radiative breaking, mass limits, and {flavor} changing neutral currents {allow} chargino, stau, stop, and
gluino NLSPs. By far, here the chargino NLSP arises most often. 
}
\label{spindepfig}
\end{figure}   
   
     Next we  note that in  the direct detection experiments,  
  the Dirac component does not give an appreciable contribution and
  essentially the entire contribution to the event rates arises from the Majoranas.
  Specifically consider the $U(1)_X$ model. Here   
 the event rates arising from the scattering of Dirac particles from nuclear  targets 
 are suppressed by a factor of $\epsilon^2$ relative to what one would find in the scattering
 of Majoranas. 
  This is easily seen as  follows: \DDM interacts with quarks in the target particles 
 by the exchange of $\gamma, Z, Z'$. The couplings of $\gamma, Z$ to  \DDM are suppressed 
 by a factor of $\epsilon$ since \DDM resides in the hidden sector.
 Thus, the cross sections arising
 from the exchange of $\gamma, Z$ are suppressed by a factor of $\epsilon^2$.  
 Next we consider the exchange of $Z'$. {The} coupling  of $Z'$ [which is mostly a $U(1)_X$ gauge
 boson] with {the Dirac component of dark matter} is assumed to be normal size, i.e.,  $O(g_X)\sim O(g_2)$. {However,} 
 its coupling with quarks is suppressed by a factor of $\epsilon$. 
 Thus the exchange of $Z'$ also
 gives a scattering cross section which is suppressed by $O(\epsilon^2)$. 
 { Since $\epsilon^2 \lesssim 10^{-4}$ {the Dirac component}
  gives a negligible number of events in the direct detection experiments relative to what the neutralinos give {as far as the $Z^{\prime}$ and $Z^{\prime\prime}$ poles are concerned}.
A concrete  analysis of events rates for the Dirac particles in these experiments has recently 
been given \cite{kctcnew} and our estimate is in accord with this analysis.  {The contribution  from the photon pole depends on the cutoff at small
angles. Further, an analysis of the event rates  is subject to absorption
both by the atmosphere as well as by dirt and rock in Earth  before the milli charged particle gets to the detector (see, e.g. \cite{Foot:2003iv}).}


\begin{table}
\begin{tabular}{ccccc}
\hline
$m_0~ \rm GeV$&$m_{1/2}~ \rm GeV$  &$A_0~ \rm GeV$  &$\delta_{1,2,3}$ & $\tan \beta$ 	\\
\hline	
676 & 148& 118 & (-0.26, 0.63, 0.53) &41\\
300 & 195& -100 & (-0.41, 0.43, 0.53) &41\\
767 & 173& -444& (-0.38, 0.60, 0.09)& 45\\
1718 & 297 & 1736 & (0, -0.37, -0.68) & 47\\
1973 & 227 & 1209& (0, -0.34, 0.34) &28\\
1152 & 139 &  1551& (0, 0.42, 0.03) & 50\\
2174 & 314 & 1537 & (0, 0.60, 0.18) &25\\

									\hline\\									
$m_{\chi^0_1 =\chi}~ \rm GeV$	&	$m_{(\tilde \chi^{\pm},\tilde g)}~ \rm GeV$	&	$m_{(\tilde \tau_1,\tilde t_1)}~ \rm GeV$	&	$m_{h}~ \rm GeV$	&	$m_{A \sim H}~ \rm GeV$	\\
\hline								
42& (168, 595)& (556, 561)& 110 & 480\\
44& (210, 723) & (211, 514) &112 &370 \\
42 &(201, 515) &(578, 508) & 111 &438 \\
120 & (134, 311) & (1233, 943)& 114 & 718 \\
86 & (103, 833) & (1822, 1259) & 116 & 1666 \\
54 & (142, 418) & (708, 639)& 110 & 475\\
106 & (144, 982) & (2042, 1402) &116 & 1920\\

\hline\\									
${(\Omega h^2)}_{\chi}$&${\langle\sigma v\rangle}^{\chi}_{\rm halo} ~\rm cm^3/s$&$\sigma(SI)_{\chi p} ~\rm cm^2$&  Ge (evts/kg/day)  & Xe (evts/kg/day) \\
   (half the DM) &  ($\chi$ halo cross sec) &  Direct Det. & ${ \rm ([10-50] KeV , {all})}$ & ${ \rm ([10-50] KeV , {all})}$\\
\hline									
$5.1\times 10^{-2}$ & $5\times 10^{-28}$ & $4\times 10^{-44}$ & $(6\times 10^{-3},1\times 10^{-2})$& $(7\times 10^{-3}, 2\times 10^{-2} )$\\
$6.5\times 10^{-2}$ & $4\times 10^{-28}$ & $3\times 10^{-44}$ & $(4\times 10^{-3},1\times 10^{-2})$& $(6\times 10^{-3},2\times 10 ^{-2} )$\\
$5.5\times 10^{-2}$ & $6\times 10^{-28}$ & $4\times 10^{-44}$ & $(6\times 10^{-3},1\times 10^{-2})$&$(8\times 10^{-3},3\times 10^{-2} )$\\
$4.6\times 10^{-2}$ & $2\times 10^{-26}$ & $3\times 10^{-44} $ & $(4\times 10^{-3}, 7\times 10^{-3} )$ & $(5\times 10^{-3},1\times 10^{-2})$\\ 
$5.1\times 10^{-2}$ & $ 2\times 10^{-26}$ & $2\times 10^{-44} $& $(4\times 10^{-3}, 7\times 10^{-3})$ & $(5\times 10^{-3},1\times 10^{-2})$\\
$5.1\times 10^{-2}$ & $3\times 10^{-28}$ & $2\times 10^{-44}$&$(3\times 10^{-3}, 6\times 10^{-3})$ & $(4\times 10^{-3},1\times 10^{-2})$\\
$5.8\times 10^{-2}$ & $3\times 10^{-26}$ & $5\times 10^{-44}$ & $(8\times 10^{-3},1\times 10^{-2})$ & $(1\times 10^{-2},2\times 10^{-2})$\\
\end{tabular}
\caption{Top {section of the} table: A sample set of NUSUGRA models which produce the  Majorana component of dark matter and 
 makes up about half the relic density of the universe for both the $U(1)_X$ and the 
 $U(1)_X\times U(1)_C$ two 
component models.
  Middle {section of the} table: Masses for light sparticles  including the  neutralino $\chi$,
  the light chargino $\chi^{\pm}$, the gluino ($\tilde g$), the light stau ($\tilde \tau_1$),  the
  light stop ($\tilde t_1$),
 and  the CP even  Higgses $h,H/A$ (charged  Higgs is slightly heavier) for the same set of inputs 
 as given in the top table. 
Bottom {section of the} table: The  Majorana relic density $(\Omega h^2)_{\chi}$, 
 {${\langle\sigma v\rangle}^{\chi}_{{\rm halo}} \ll {\langle\sigma v\rangle}^{\psi}_{{\rm halo}}$}, spin-independent neutralino proton cross section $\sigma(SI)_{\chi p}$,
 and event rates/kg/day in {germanium} and {xenon} detectors corresponding to entries in the top table with an assumed  {30\%} efficiency 
 and with $\rho_{\odot,\chi} \sim (1/2)\rho_{\odot,\rm total} \sim 0.18$ ${\rm GeV/cm^3}$. 
 The analysis is done with a  top pole mass at 171 GeV and the models show stability in the relic density with small changes in the pole mass. 
 The $\sigma(SI)$ given here are also within the range of XENON-100 and are on the edge of the 
 limits recently reported in Ref. \cite{Aprile:2010um}. More low mass models can be seen in Fig.(\ref{spindepfig}).
}
\label{tabbenchers}
\end{table}

 For the $U(1)_X\times U(1)_C$ model, the Dirac particles have no interaction
with the quarks, {so} the contribution of the Dirac particles to event rates in the direct  detection 
experiments is absent. Thus in either case the dominant contribution to event rates in experiments
for the direct detection of dark matter comes from {neutralinos.} 
In 
Fig.(\ref{spindepfig}) we give an analysis of {spin-independent} cross section in NUSUGRA models 
for the parameter space
of supergravity models with nonuniversalities in the gaugino sector so that $m_{a}=m_{1/2}(1+\delta_a)$
{with $(a=1,2,3)$}.
In the analysis the NUSUGRA parameters are chosen in the following range:
 $m_0 < 3 ~\rm  TeV$, $m_{1/2}  < 400 ~\rm GeV$, $\delta_{a=2,3}$ lie in the range $(-1,1)$,
 (which statistically {favors} the low LSP mass region {which  is also the region of interest in 
 this analysis})  
	 $|A_0/m_0|<4 $,  and {$\tan \beta = (1-60)$}. The current limits from  CDMS,  XENON, and
	from  other experiments are also exhibited. We are 
{assuming the neutralinos are contributing roughly
half the relic abundance  and  roughly half the local density of dark matter}. There is no rescaling
 by the dark matter density in these figures. 
Note the models are dominated by chargino NLSPs  \cite{msp,Berger} and the presence of a low mass chargino wall  \cite{Feldman:2007fq}.
The relatively empty region in the range of (70-90) GeV follows from the constraint on the chargino mass being larger than 100 GeV, and
an inability  for the chargino and LSP to therefore coannihilate in this region, along with the constraints that
the stop and stau are larger in mass than 100 GeV, and the gluino should be larger than about 300 GeV. 
{Further,}
the lightest CP even
Higgs has been constrained to lie higher than 110 GeV.  The region
of low mass is mostly controlled by the poles of the MSSM, while the higher mass region above 100 GeV is controlled mostly
by coannihilations.

\section{{Collider Signals}}

We discuss now the collider implications of the $U(1)_X$ and {the} $U(1)_X\times U(1)_C$  models.
 In Table(\ref{tabbenchers}) we give some concrete models which generate half the relic abundance
from the neutralino dark matter. These {models} produce event rates in {germanium} and in {xenon} at 
detectable {levels with a relatively light spectrum}.
The predicted spin-independent elastic WIMP-nucleon cross {sections}
are on the edge of the limits reported by XENON-100 \cite{Aprile:2010um}.
Specifically, all of the models listed  in Table \ref{tabbenchers} have a  light neutralino
mass  and several also have light {Higgses} (for recent work 
relating to light {Higgses} and the COGENT  \cite{Aalseth:2010vx} and DAMA  \cite{Bernabei:2008yi}
data, see  \cite{Feldman:2010ke},  \cite{Kuflik:2010ah} and   \cite{Andreas,Bae:2010ai}). 
Further, essentially all the models in the table have a gluino lying in the {sub-TeV} range 
and typically all the charginos are light.  However, some of {the} models have rather {large scalar
masses in the (1.5 - 2) TeV} region indicating that they originate on the {hyperbolic branch} of radiative
breaking of the electroweak symmetry \cite{ccn}.  
The {models} listed  in Table \ref{tabbenchers} share the 
property that the gauginos in all cases are relatively light. Thus, such models 
 should give rise to detectable signals in the form of leptons and jets and missing energy at the LHC
 with modest luminosity {(though the missing energy may be difficult to estimate in early runs)}. 
{The models with very light gluinos could surface with less than 1 {\rm fb$^{-1}$} {at LHC center 
of mass energies} $\sqrt s = 7,10 ~\rm TeV$,(for recent analyses see 
 \cite{Feldman:2009zc}, \cite{FKRN} and  \cite{Izaguirre:2010nj}) while many of the models should be discoverable with  O(10)~fb$^{-1}$
{at} $\sqrt s = 10 ~\rm TeV$.  {(In fact one can glean this from the analysis of the first {listing of Ref.}  \cite{flncdms}.)}}
As many of the candidate models have rather light gluinos with a chargino NLSP,  such models likely will produce
missing energy which is very SM like.  
{Large event rates can arise, however, from multijets, and,  in particular,  from  b jets.}
One also expects a {sizable} amount of leptons in these models. For the cases where 
{stau-coannihilation} survives, the {leptonic signals are} likely to be stronger and the missing energy larger than for the chargino NLSP cases. However,
since these models have very low SUSY scales, most of them should indeed be discoverable (and likely rather early)
at the LHC.
Additionally for the $U(1)_{X}\times U(1)_C$ model $Z^{\prime \prime}$ offers the possibility of discovery at 
an NLC {as it will have distinct signatures}. Its decay width is significantly smaller than what a {GUT-type}
$Z'$ with the same mass will have. Additionall, it has visible decays only into  $e\bar e$ and $\mu\bar \mu$
{along with  radiation at the NLC}. 
Thus, a $Z^{\prime \prime}$ of this type can be  detectable at an NLC because of its distinct signatures.
However, a full simulation of collider signals {requires a separate dedicated analysis.}
\\


 \section{Conclusion}{\label{conclusions}}
 {In this work} we  have proposed a new class of models  
 with dark matter consisting of two, three, or even four components.  We considered the two component
 model consisting of Dirac and  Majorana  particles in detail. We showed that this
{two component} 
  model can fit the positron excess seen in the PAMELA experiment 
 as well as can produce detectable signals in the current direct detection experiments while satisfying 
 WMAP relic density constraints.   Thus. the   Dirac  component of  the two component dark matter 
 {model}
  allows  a fit to the PAMELA data via annihilation of the Dirac particles close
 to a Breit-Wigner pole. 
 On the other hand,  the Majorana component of dark matter 
plays the dominant role in the generation of events in dark matter detectors. Specifically,
we  showed that in the two component picture it is possible to generate events of size 
1-2 in 612 Kg-d of data in the {CDMS-II} detector as well as event rates that can be 
tested by the results of XENON-100, and an observable number of events in other 
ongoing direct detection experiments.
Further, it was shown that models which lead to detectable signals in direct detection experiments 
 are typically associated  with a relatively light spectrum
which is discoverable at the LHC with modest luminosity. 
 Further, one class of models discussed in this
work produces  a $Z^{\prime \prime}$ vector boson which has visible decays only to $e^+e^-$ and $\mu^+\mu^-$.
The proposed models contains massive  scalar fields which are also possible candidates for dark matter.
 {Thus these spin zero fields in combinations with Dirac and Majorana particles  present
  the possibility of a multicomponent dark matter.} 
 Finally, we note that it would be very interesting to investigate phenomena where both components
  play a significant role, i.e., regions of the parameter space of the model which allow both dark matter candidates
  to appear in the data analysis.  Such a possibility may appear in certain decay fragments at the LHC
  where the missing energy signals from the two dark matter particles would be different because
  of their different masses and interactions. \\

\noindent
{\em {Acknowledgements}}:  DF thanks Kenji Kadota  for a communication.
This research is  supported in part by DOE {Grant No.}  DE-FG02-95ER40899
 (Michigan - MCTP), NSF {Grant No.} PHY-0757959 {(Northeastern)},
and NSF {Grant No.} PHY-0653342 (Stony Brook - YITP).



\end{document}